\documentclass[draft,jgrga]{agu2001}
\usepackage{graphicx}
\authorrunninghead{Srivastava et al.}

\titlerunninghead{Source region of CME of November 18, 2003}

\authoraddr{Nandita Srivastava,
Udaipur Solar Observatory, Physical Research Laboratory, Udaipur.
(nandita@prl.res.in)}

\begin{document}

%
%
%
%
%

%
%

\title{Source region of the 2003 November 18 CME that led to the strongest magnetic storm of cycle 23}
%

%
%


\author{Nandita Srivastava, Shibu K. Mathew, Rohan E. Louis}
\affil{Udaipur Solar Observatory, Physical Research Laboratory, Udaipur-313001, India}

\author{Thomas Wiegelmann}
\affil{Max-Planck-Institut f\"ur Sonnensystemforschung,
Max-Planck-Strasse 2, 37191 Katlenburg-Lindau, Germany}

\begin{abstract}
The super-storm of November 20, 2003 was associated with a high
speed coronal mass ejection which originated in the NOAA AR 10501
on November 18. This coronal mass ejection had severe terrestrial
consequences leading to a geomagnetic storm with D$_{ST}$ index of
-472 nT, the strongest of the current solar cycle. In this paper,
we attempt to understand the factors that led to the coronal mass
ejection on November 18. We have also studied the evolution of the
photospheric magnetic field of NOAA AR 10501, the source region of this coronal mass ejection. For this purpose,
the MDI line-of-sight magnetograms and vector magnetograms from
Solar Flare Telescope, Mitaka, obtained during November, 17-19,
2003 were analysed. In particular, quantitative estimates of the temporal
variation in magnetic flux, energy and magnetic field gradient were
estimated for the source active region. The evolution of these
quantities was studied for the 3-day period with an objective to
understand the pre-flare configuration leading up to the moderate
flare which was associated with the geo-effective coronal mass
ejection. We also examined the chromospheric images recorded in
H$_\alpha$ from Udaipur Solar Observatory to compare the flare
location with regions of different magnetic field and energy. Our
observations provide evidence that the flare associated with the
CME occurred at a location marked by high magnetic field gradient which led to release of free energy stored in the active region.
\end{abstract}

%
%

%

\begin{article}

%
%

\section{Introduction}
One of the major challenges in space weather prediction is to
estimate the magnitude of the geomagnetic storm based on solar
inputs, mainly the properties of the source active regions from
which the coronal mass ejections ensue (Srivastava 2005a, 2006).
Several specific properties of the source regions of CMEs and their
corresponding active regions have been studied by various authors,
for example,  speeds of the halo CME (Srivastava and
Venkatakrishnan, 2002, 2004; Schwenn et al. 2005), source active
region energy and their relation to speed of the CMEs
(Venkatakrishnan and Ravindra 2003; Gopalswamy et al. 2005a).
These studies are aimed at understanding the solar sources of
geo-effective CMEs and using this knowledge in developing a
reliable working prediction scheme for forecasting geomagnetic
storms (Schwenn et al. 2005, Srivastava 2005b, 2006). It is
important to point out that continuous  observations made
available with the launch of SoHO suggest that most of the major
geomagnetic storms (with D$_{ST} \leq -100$ nT)  are accompanied by high speed halo CMEs which,
in turn, are associated with strong X-class flares.  For example,
Srivastava (2005a) found that geomagnetic storms with D$_{ST}$
$\sim -300$ nT are related to strong X-class solar flares
originating from low latitudes and located longitudinally close to
the center of the Sun. These studies assume importance as
properties of the source active regions could form the basis of
solar inputs for developing a  predictive model for forecasting space weather.

The motivation of this study stems from the observation that the
source active region did not exhibit any  solar characteristics
significant enough to render the intense storm on 20 November
2003. This is  an exception from the other super-storms (D$_{ST}$
$\sim -300$ nT) of the current solar cycle described by Srivastava
(2005a) and Gopalswamy et al. (2005a). This event is therefore,
significant from the perspective of space weather prediction and
requires a detailed investigation in order to understand the
factors leading to such an event. Although most super storms
studied by Srivastava (2005a) were associated with high speed CMEs
and strong X-class flares in large active regions, the most
intense storm of 20 November 2003 (D$_{ST} \sim -472$ nT) had its
source in a relatively smaller and weaker M3.9 class flare. This
posed a real challenge for the space weather forecasters as the
source of this geomagnetic storm was a CME with a moderate
plane-of-sky speed of $\sim 1660$ km s$^{-1}$. Detailed studies on
the CME of November 18, 2003 CME made by Gopalswamy et al. (2005a)
and Yurchyshyn et al. (2005) reveal that the geomagnetic storm
owes its large magnitude to the high interplanetary magnetic field
(52 nT), strong southward component of the interplanetary magnetic
field (-56 nT) and the high inclination of the magnetic cloud to
the plane of the ecliptic which ensured a strong magnetic
reconnection of the magnetic cloud with the earth's magnetic
field. This also enhanced the duration for which solar wind-magnetospheric interaction took place which was 13 hours as
against a few hours for even the super-storms with D$_{ST}$ index
(-300 nT) recorded in the current solar cycle (Srivastava 2005a).
The question is: what triggered this eruption of magnetic cloud
from the Sun. In order to answer this question, we investigated
the properties of the source active region NOAA AR 10501 of the
November 18, 2003 CME. We compared the pre-flare/CME and the
post-flare/CME magnetic field configuration and also  studied the
variation in the magnetic field gradient and the available
magnetic energy in the source active region.

\section{Observational Data}
The present study on the source active region of the CME is based
on (i) H$_{\alpha}$ filtergrams from the Udaipur Solar
Observatory, India; (ii) line--of--sight magnetograms obtained
from the Michelson Doppler Imager (MDI) instrument aboard SoHO
spacecraft (Scherrer et al. 1995) (iii) vector magnetograms from
the Solar Flare Telescope (SFT), at Mitaka, Japan (Sakurai et al.
1995) and (iv) associated white light CME data from Large Angle
Spectrometric coronagraph (LASCO) (Brueckner et al. 1995).

The H$_{\alpha}$ chromospheric filtergrams used in this study were
obtained during 5:00 to 10:00 UT on November 18. The image cadence
in H$_{\alpha}$ varies from a few frames per minute to one frame
per minute for the period of study. The spatial sampling of the
H$_{\alpha}$ filtergrams is approximately 0.6 arc-sec per pixel
with a field of view of $752 \times 480$ pixels.  Full disk
line--of–-sight magnetograms were obtained for the period
November, 17 to 19, 2003 from the MDI instrument aboard SoHO.
These images are available at a cadence of one minute and have a
spatial sampling of $1.98$ arc-sec per pixel with a field of view
of $1024 \times 1024$ pixels.

We also used the small field high resolution vector magnetic field data
for the same active region for 17 and 18 November (one image per day)
recorded by the Solar Flare Telescope at Mitaka National Astronomical
Observatory of Japan (http://solarwww.mtk.nao.ac.jp/en/database.html). This
instrument measures the photospheric vector magnetic field using Fe 6302.5
\AA\ line. The field of view of the vector magnetogram is $256 \times 240$ pixels, where 1
pixel $= 0.66$ arc-sec.

Observations from GOES X-ray satellite showed an M3.9 class flare starting
 in the active region NOAA AR 10501  at 8:00 UT and attaining peak intensity
at 8:30 UT. Figure 1 shows the time-lapse H$_{\alpha}$ images of the flare which
started at a location close to the southern sunspots, and
spread along the neutral line assuming the shape of a classic two-ribbon
flare. The southern portion of the circular filament was
blown off at 7:53 UT, which coincides with the timing of the launch of the associated CME.
As a matter of fact, two CMEs were recorded in this active region on 18 November, at
8:06 UT and 08:50 UT. The first CME was associated with an M3.2 flare and was confined mostly to the southeast with minimal overlap with the earth direction, therefore the magnetic cloud of November 20 was identified to be associated with the the second CME.  This eventually 
led to the strongest geomagnetic storm at the earth (Gopalswamy 2005b and Srivastava 2005a). In fact there were other CMEs on November 19 from the same region but they were too slow to be considered as the source of the observed magnetic cloud.

\section{Data Analysis and Discussion}

We analysed a series of magnetograms  taken at 1 minute cadence
during November $17-19$, 2003. For the analysis, we selected 168 images
spanning the above period with an interval of 15 minutes. The bad images in the data-set were replaced by the ones closest in time. The images were first corrected for the solar rotation taking
the last image on 18 November as the reference and then registered using a
two-dimensional cross-correlation program. From these full disk magnetograms, a smaller region covering the AR 10501 including the filament
channel was extracted for analysis.

Figure 2 shows an overlay of the magnetic field contours obtained
from one of the magnetograms taken at 06:24 UT on an H$_{\alpha}$
image taken around the same time. The H$_{\alpha}$ image and the
magnetogram were co-aligned to contain the same area of interest.
These co-aligned images were used to identify some of the
sub-areas within the active region where most of the changes in
magnetic flux or the initial flare activity in H$_{\alpha}$ images
were observed. The sub-areas were chosen as a box of 1 arc-min
size to accomodate the sunspot as well as its neighbouring region
to detect any anomalous changes in the above mentioned
quantities that might have led to flare. Considering the size of
the sunspots, an optimum size of the box was chosen to include its
neighbouring region as well. Choosing a larger box size would make
it difficult to ascertain which region was responsible for the
changes in the aforementioned parameters and a  small box size
would lead to a contamination by both emergence of flux as well
motion of magnetic inhomogeneities, a consequence of the poor
resolution of the magnetograms. The sub-regions are marked with boxes in Figure 3. The fact that there is no correlation in the flux in the two selected boxes 1 and 2 (cf. Figure 4) implies that variation of flux is consistent for the size of the box although a threshold value was not set as stated  by  Lara et al. (2000).
It may be noted here that the source active region contains sunspots which are extremely complex as both the main spots have umbrae of opposite polarities  within the same penumbra. Further, the initiation of the flare took place close to the umbrae of the sunspots in the box, labeled as  '1'. 
 Using the
magnetograms, a number of parameters were estimated for this
active region such as magnetic potential energy, magnetic flux and
magnetic field gradient.

The  magnetic energy for a potential field configuration was computed for the active region using the virial theorem (Wheatland and Metcalf, 2006; Metcalf et al. 2008)
\begin{eqnarray}
 E_p = \frac {1} {4 \pi} \int (xB_{px}+yB_{py}) B_z dx dy
\end{eqnarray}

where E$_p$ is the available potential energy in the region of
interest. The origin of the coordinate system here is taken to be
the center of the region of interest. The photospheric magnetic field components, B$_{px}$ and B$_{py}$  in x and y directions have been computed under the assumption of a potential magnetic field using Fast Fourier Transform (FFT) described by Alissandrakis (1981). These
potential field components were then used to compute the magnetic
energy using Equation 1. The parameters x and y signify the
distance on the Sun having a transverse field B$_x$ and B$_y$
respectively.

Further, we estimated the magnetic flux in the
active region using
\begin{eqnarray}
\int {B_z da}
\end{eqnarray}

for the positive and negative polarities of the active region separately, where
{\it da} is the elemental area. The magnetic field gradient in the active region
can be computed using the following equation:
\begin{eqnarray}
\nabla B_{z} = \sqrt{\left (\frac{dB_{z}}{dx} \right )^2  + \left (\frac{dB_{z}}{dy} \right )^2}
\end{eqnarray}


\subsection{Magnetic flux variation in the AR 10501}
The SoHO/MDI magnetograms measure the line--of--sight component of the
magnetic field, B$_z$. The area within which the flux is calculated
is the selected area of the active region in each pixel. This was determined
using the image scale and the angular scale of the Sun from the image header,
which corresponds to $1409 \times 1409$ km$^{2}$. The magnetic flux of
positive and negative polarity was computed separately for each sub-region
marked by boxes in Figure 3. Then, the variation of magnetic flux with time was studied
(Figure 4). The figure shows that  the  negative flux for both the regions '1' and '2' and the positive flux in the region '1' show an increase with time, until
8:30 UT on November 18. This time coincides with the time of the M3.9
flare/halo CME on this day. After 8:30 UT, the magnetic flux values decreased.

In the region 1, the negative flux  increased from 2.5 to 3.0
$\times10^{21}$ Mx and then decreased  to a value of 1.62$\times
10^{21}$ Mx after the flare/CME on  November 18.  The positive
flux increased from $1.3 \times 10^{21}$ Mx to $ 1.8\times
10^{21}$ Mx. On the other hand,  region 2 showed an increase in
the negative flux from 4.8 $\times 10^{21}$ Mx to 5.5 $\times
10^{21}$ Mx. This region  shows negligible variation in the
positive flux.  The positive and negative flux in the
region 3 show no variation, as expected, since the noise in the magnetogram is of the order of $\pm10$ G (Scherrer et al. 1995).

Although the magnitude of the negative flux is higher (almost twice) in
region 2 than in region 1 the rate of the increase of negative flux in both the
regions is approximately the same, ($\sim 5.5 \times 10^{15}$
Mx s$^{-1}$). It is  found that in region 1, the positive flux also
increased slowly with time, with absolute values higher than those of
region 2. Thus, in region 1, the total flux increase is due to the increase in both
the fluxes; while in region 2, the total flux increase is entirely due to the increase in the negative polarity flux. 
An overall increase in the absolute flux until the time of flare indicates that the emergence of new flux in the active region might
have played a key role in triggering this flare/CME, particularly,
in  region 1. These indicate that the initiation of the flare is well correlated with the evolution of flux. 

\subsection{Variation of magnetic field gradient}
We also estimated the value
of average magnetic field gradient for the three small regions marked in Figure 5.
 Our measurements showed that  the average gradient peaked to
$\sim 90$ G Mm$^{-1}$ just before the flare in region 1.
Region 2 also shows a sharp rise in the average gradient to  84 G
Mm $^{-1}$ before the flare. While there is a conspicuous rise in the average
field gradient in both the regions, there are minor peaks in between
that are possibly related to several other minor flares/CMEs which were launched from
the same active region. An overlay of the gradient of magnetic field shows that the maximum gradient $\sim
90$ G Mm$^{-1}$ and occurred at the location where the flare kernel first appeared in H$_\alpha$. It is to be noted that the flux and gradient seem to be well correlated. The plots also indicates that the initiation of the flare is well correlated with the magnetic field gradient in the region it occurred.

\subsection{Variation in the magnetic energy}
The magnetic potential energy calculated for the three small regions marked
as 1, 2 and 3 show that the magnetic energy is the highest for
the region 2 and of the order of $10^{32}$ ergs (Figure 6). For the regions 1 and 3, the magnetic energy is
smaller of the order of $10^{31}$ and $10^{30}$ ergs respectively. One of the explanation for the large values of the magnetic
energy in  region 2 is that it includes a full big sunspot, which entails high magnetic flux and hence higher magnetic energy.

We measured the magnetic energy of the entire NOAA AR10501 for
comparison, using the code developed by Wiegelmann (2004) to extrapolate the coronal
magnetic field lines with vector magnetograms as input. These magnetograms were pre-processed with the help of
a minimization method as described in Wiegelmann et al (2006).
This method is superior to both the potential field extrapolation
model and the linear force-free field extrapolation models as
shown by Wiegelmann et al. (2005). The former has been used by
several authors to compute the potential energy, because of its
simplicity, (Forbes 2000, Venkatakrishnan and Ravindra, 2003,
Gopalswamy et al. 2005a). However, it has been found that both
these models are too simplistic to estimate the magnetic energy
and the magnetic topology accurately (Schrijver et al. 2006 and
Wiegelmann, 2008 and references therein). We compared the
potential field energy of the entire active region NOAA 10501 on
17 (01:44 UT) and 18 November 2003 (00:20 UT) using the vector
magnetogram data obtained from the Solar Flare Telescope at
Mitaka. The entire computational box of 256x240x200 pixels size was chosen. Here, 1 pixel has a size of 1.32 arc-sec. The extrapolated field lines have been plotted in Figure
7. It is obvious from the plot that the field lines are highly
twisted and non-potential close to region 1, and over the
neighbouring curved filament in the active region which eventually
erupted. The calculation shows that the magnetic energy over the
entire active region is of the same order as that of region 2. The
errors in the magnetic energy calculation using Wiegelmann (2004)
code has been estimated to be within 3-4\% for cases where the
majority of magnetic flux is located sufficiently far away from
the lateral boundaries of computational box and about 34\% if high
magnetic flux occur close to the boundaries (Schrijver et al.
2006; Wiegelmann et al. 2006, 2008). For the  active
region for which the magnetic energy has been calculated in this paper, the
majority of the flux is located far from lateral boundaries of the computational box and 
magnetic strength flux is low close to the side boundaries. Therefore, one can
consider an error of less than 5\% in the estimated magnetic
energy.

\begin{table}
\hspace*{0.5cm}
\label{table1}
\caption {Estimates of the energy for the NOAA AR10501}
\hspace*{0.5cm}
\begin{tabular}{llll}     
\hline
Date &  Potential field &  NLFF &Max Free \\
Time  & Energy (ergs) & Energy (ergs) & Energy (ergs)\\
\hline
 17 Nov 2003  & $4.07 (\pm 0.20) \times 10^{32}$ & $6.72 (\pm 0.34)\times 10^{32}$& $2.65 (\pm 0.13) \times 10^{32}$\\
01:44 UT & &\\
 18 Nov 2003  & $3.3 (\pm 0.17)\times 10^{32}$ & $5.65 (\pm 0.28)\times 10^{32}$& $2.35 (\pm 0.12) \times 10^{32}$\\
00:20 UT &
 &\\
\hline
\end{tabular}
\end{table}

From Table 1, it is evident that the magnetic potential  energy arising from a
nonlinear force free field in the Active Region NOAA 10501 was higher on
November 17, 2003 than on November 18, 2003, even before the CME
took off. This can be reasonably explained by the fact that the same
active region triggered off another CME the previous day i.e. 17 November at
8:57 UT. This CME was observed as a partial halo CME recorded by LASCO coronagraphs and
was associated with an M4.2, 1N class flare. However, because of lack of co-temporal vector
magnetograms, it is difficult to confirm this
explanation.

Further, it is well established that it is the magnetic free
energy in an active region that powers an ensuing coronal mass
ejection and that the maximum speed of the coronal mass ejection
is constrained by the maximum free energy available in the source
region. As pointed out by Metcalf et al. (1995), free energy in
the magnetic fields can be  estimated from the distribution of the
current in the coronal layers. Table 1 also shows that the maximum
free energy available on  November 17 and 18 is respectively,
about $0.65 (\pm 0.0325)$ and $0.7 (\pm 0.035)$ times that of the
corresponding estimated potential energy. Here, the errors in the
computed magnetic energy has been estimated to be less than $5\%$
(Schrijver et al. 2006, Wiegelmann et al. 2008). This suggests that
the assumption that the available potential energy may be a good
indicator of the free energy may not always be true, as has been
assumed by several authors viz. Venkatakrishnan and Ravindra
(2003), Gopalswamy et al. (2005a) in the absence of vector
magnetograms. For better estimates of the available free energy,
vector magnetograms taken at a higher cadence are required.

\subsection{Magnetic energy and CME speeds}

The  maximum projected plane-of-sky speed of the CMEs of  November 17 and 18 (as estimated  from the
LASCO/SoHO coronagraph data) were of the order of 1000 km s$^{-1}$ and 1660 km s$^{-1}$, respectively. It is important to
mention here that the flare classification in X-ray for the November 17 flare is M4.2,
which is relatively higher than the M3.9 for the November 18 flare. Further, the free energy
available on November 17  is higher  compared to that on November  18, but the CME speed is
higher for the latter.

We obtained the mass of the CMEs of November 17 and 18, 2003 (Vourlidas 2009, {\it Private Communication}). We then computed the
kinetic energy of the CMEs under study using the measured values
of the speeds of the CMEs.  The estimated maximum kinetic energy for the
CME of  November 17 is about $0.3 \times 10^{32}$ ergs. It may be
pointed out here that the kinetic energy estimate can be uncertain
by a factor of 2 because of the uncertainty involved in estimation
of the CME mass (Vourlidas et al. 2000). However, even with this uncertainty, kinetic
energy is only a fraction of the maximum free energy, suggesting
that only a  fraction of the maximum free energy was spent in
launching the CME.  On the other hand, we found that the estimated
kinetic energy of the CME on the November 18 is $3.3 \times
10^{32}$ ergs, a value higher than the available free energy. the
reason for this discrepancy, may be well due to  the uncertainties
involved in measurement of the kinetic energy. 

As mentioned above, due to uncertainty  in the measurement of mass, the measured kinetic energy is  uncertain by a factor of 2. Taking this into account we find that the kinetic energy of the CME on November 17, CME  can vary from $1.6 \times 10^{31}$  to $6.4 \times  10^{31}$  ergs, which approximately corresponds to 5.5\% and 22\% of the available free energy. Although the observations are not co-temporal as in the case of November 18 CME, the estimated kinetic energy with the given uncertainty is still less than the estimated free energy. 

If we extend the same argument to the CME of November 18, it is observed that the uncertainty in the estimated kinetic energy varies from $1.65 \times 10^{32}$   to $6.6 \times 10^{32}$  ergs. If one assumes the former value, the kinetic energy is approximately 70\% of the available free energy while the latter exceeds the available free energy by a factor of 2.8. Since we are limited by lack of simultaneous observations, it would be inappropriate to quantify the small yet finite difference in available free energy in the active region and the estimated kinetic energy of the CME on November 18 CME originating from same active region.

Another possibility
for this discrepancy could be the fact that  the free energy on
November 18 was calculated for the time at which the vector
magnetogram was available, which was eight hours before the CME
was actually launched. There is a possibility that the free energy
was lower at this instant and  had since risen. This is
supported by the argument that the plot of the magnetic potential
energy derived from the line-of-sight MDI magnetograms shows a
rise during this phase. It is more likely that the total energy is
also large owing to increase in magnetic flux. This also
underscores the importance of obtaining regular vector
magnetograms  at a higher cadence. A study of source regions of
geo-effective CMEs in this cycle using Hinode vector magnetograph
observations may be extremely helpful to resolve similar issues.

\section{Summary and Conclusion}
The analysis of the magnetic field data of the source active region of November 18, 2003 CME, before, during and after the
flare, lead to the following inferences.

Of the three regions, region 2 possesses the largest magnetic
energy, and magnetic flux. It also shows a steeper rise in the
magnetic field gradient than the other two regions. This indicates
that initiation of the flare may occur at this region. However,
the flare in H$_\alpha$ initiated at a location that is marked by
high average gradient and the emergence of fluxes of both
polarities. In fact, the rate of increase of magnetic flux is the
same for both the regions 1 and 2. Moreover, the region associated
with the flare/CME onset is also marked by twisted non-potential
low-lying field lines, while region 2 is marked by straight
non-twisted open field lines, as evident from the field line
extrapolation. It is to be kept in mind that here the definition of open field lines here signify the field lines which pass through the upper boundary (200 pixel or  264 arc-sec). These 'open field lines' do not close within the active region. It is not possible to distinguish, if they are globally open or connected to areas outside the considered active region.

This indicates that the best configuration for
reconnection may have occurred in region 1, because new fluxes of
both polarities emerge and maintain a high magnetic gradient in
the small region chosen for the analysis. Once the flare is
initiated and spreads in the shape of two-ribbons the field lines
then get reconnected to the twisted magnetic field lines over the
filament, leading to its destabilization and eruption as a whole.
The extrapolation of non-linear 3-D force free field lines above
the active region is shown in Figure 7. This clearly shows the
twisted field lines above the filament which erupted with the
flare and associated CME.

\begin{enumerate}

\item The pre-flare configuration of the active region NOAA AR 10501 is marked
by emergence of flux in both polarities thereby increasing the total flux and
high magnetic gradient of the order of 90 Mx s$^{-1}$ at a localized site of
flare initiation.

\item The time of the initiation of the flare is well correlated
with the evolution of the flux and gradient in the region it first
appeared.

\item The nonlinear force-free field line extrapolation shows that the
region of the flare/CME onset is marked by twisted non-potential low-lying
field lines as compared to the other region, which is marked by straight open
field lines as evident from the field line extrapolation. The flare triggered
the reconnection of the field lines overlying the neighbouring filament in
the active region which is also highly non-potential in nature.

\item The total magnetic potential energy of the active region was estimated to
be of the order of 10$^{32}$ ergs. The magnetic energy of the active region
increased continuously before the flare.

\item The maximum free energy available in the active region is  approximately 0.7 times
that of the potential energy.

\end{enumerate}

%
%

\begin{acknowledgments}
The authors would like to acknowledge the MDI/SoHO team for the magnetic
field data used in this paper. We also thank the observers at USO for
obtaining the H$_{\alpha}$ data used in this paper. The authors thank the
LASCO/SoHO consortia for their data on the CME. SoHO is an international
cooperation between ESA and NASA. The authors also express their thanks to
Prof. T. Sakurai for providing the vector magnetograms obtained by the Solar
Flare Telescope site at Mitaka, Japan for this study. A part of this work was
done by NS at the Max-Planck-Institut f\"ur Sonnensystemforschung, Germany.
She acknowledges the institute for the financial support for her visit. TW was supported by DLR-grant 50 OC 0501.
\end{acknowledgments}

\begin{figure*}
\includegraphics[width=16cm]{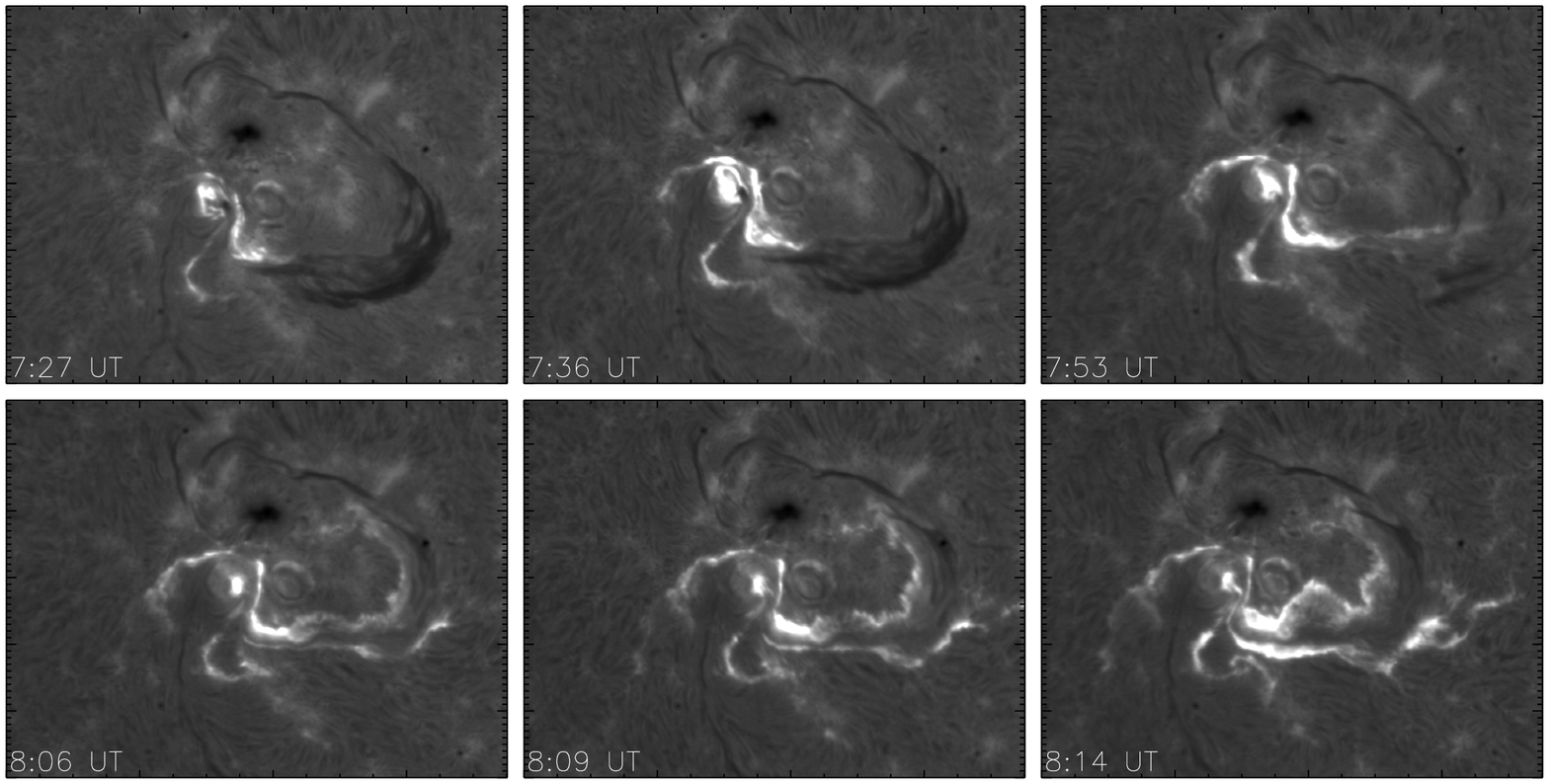}
\caption{The M3.9, 2N flare observed in H$_{\alpha}$ at Udaipur Solar
Observatory on  November 18, 2003 in NOAA AR 10501. } \label{fig1}
\end{figure*}

\begin{figure}
\includegraphics[width=12cm]{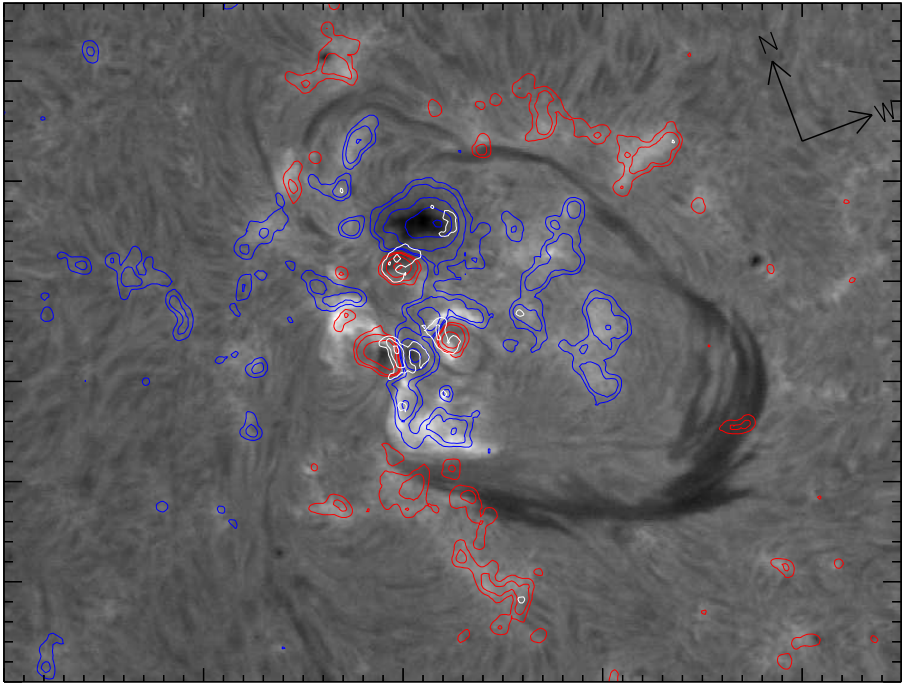}
\vspace{0.3in}
 \caption{The magnetic field contours obtained from
the line-of-sight MDI magnetogram, overlaid on  H$_\alpha$ image
obtained from the Udaipur Solar Observatory. Both the images were
recorded at 6:24 UT. The red and blue contours here denote the
positive and negative polarities respectively with field strength
values of ± 1500, 1000, 200 and 100 G. The white contours
overplotted on this image are the locations of strong magnetic
field gradient derived from the line-of-sight magnetogram recorded
at the same time. The two boxes correspond to the sub-areas
considered for the study of magnetic flux evolution in MDI
magnetograms shown in Figure 3.} \label{fig2}
\end{figure}

\begin{figure}
\includegraphics[width=15cm]{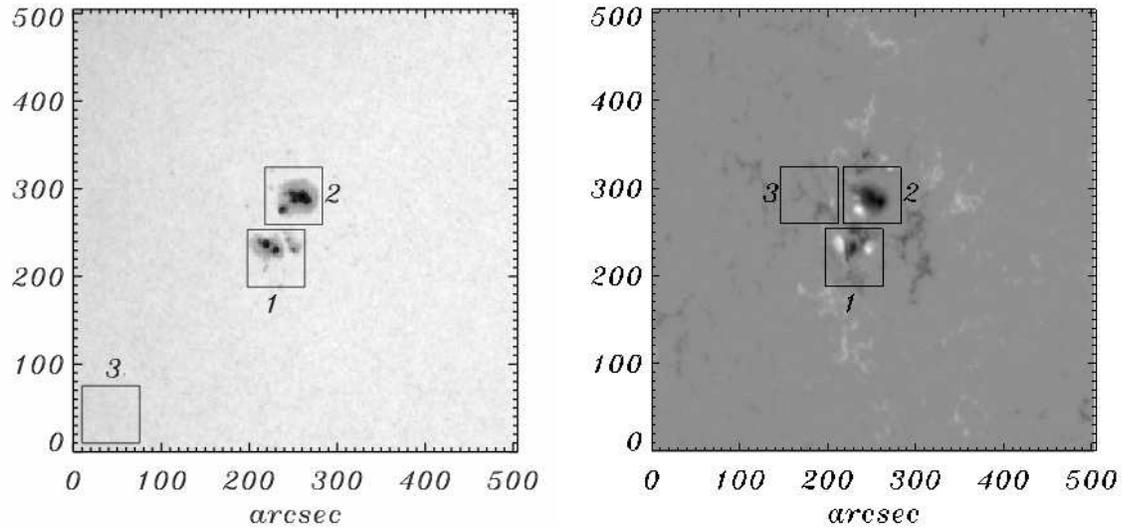}
\caption{The left and right panels show the MDI continuum image
and magnetogram taken on 18 November, respectively for the active
region NOAA 10501. We studied the evolution of magnetic flux,
average gradients in this region at 3 different locations marked
by square boxes and named as 1,2,3. In order to compare, we chose
these three regions to include northern sunspot of the AR in Box
2, southern sunspot in Box `1' and relatively quiet region in Box
3. These three regions were $67 \times 67$ arc-sec$^2$ size with the
box size of $34 \times 34$ pixels with each pixel corresponding to
1.98 arc-sec.} 
\label{fig3}
\end{figure}

\begin{figure}
\includegraphics[width=9.5cm, angle=90]{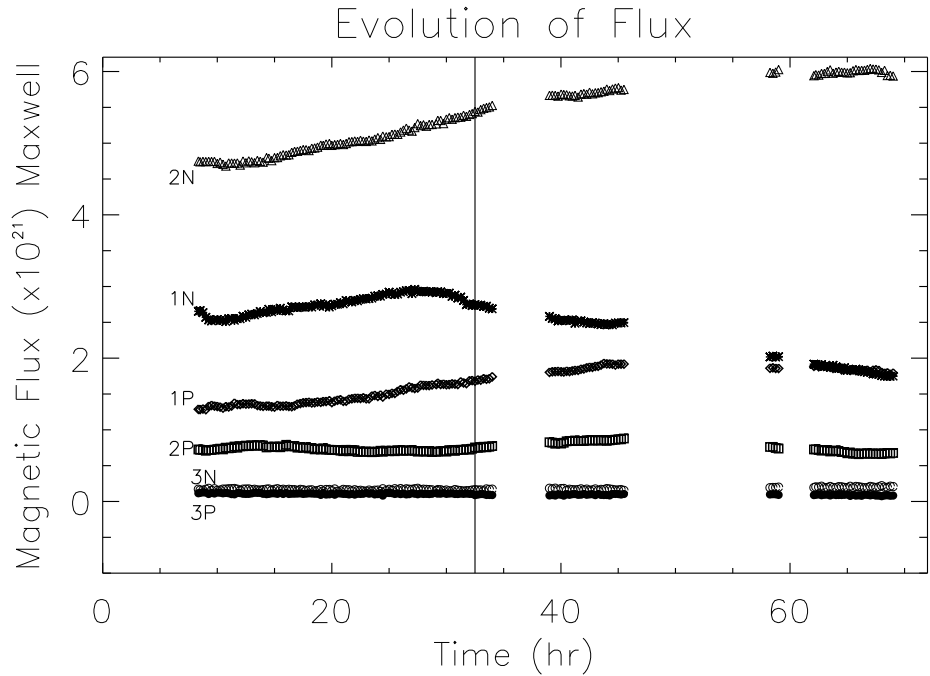}
\caption{This figure shows  evolution of the positive and negative
magnetic flux in the separate boxes marked by 1,2,3 in Figure 3.
The positive and negative flux for the three regions are  shown as asterix, diamond, square box, triangle, open circles and filled circles respectively. The x-axis indicates time with start time as 0 UT on 17
November. The vertical line at 32.5 hour coincides with the peak of
the flare. The discontinuity in the plots indicates a gap in the
data recorded by the MDI instrument.} \label{fig4}
\end{figure}

\begin{figure}
\includegraphics[width=9.5 cm, angle=90]{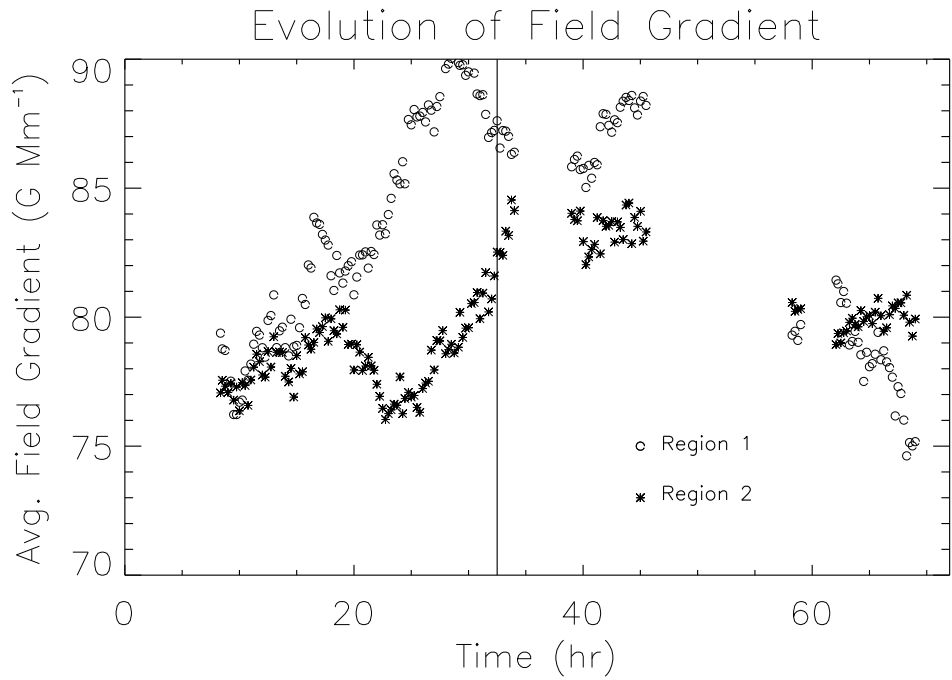}
\caption{Evolution of magnetic field gradient for the  regions 1 and 2 in Gauss
per Mm. The plots show an increase in the magnetic field gradient for both the 
regions.}
\label{fig5}
\end{figure}

\begin{figure}
\includegraphics[width=9.5cm, angle=90]{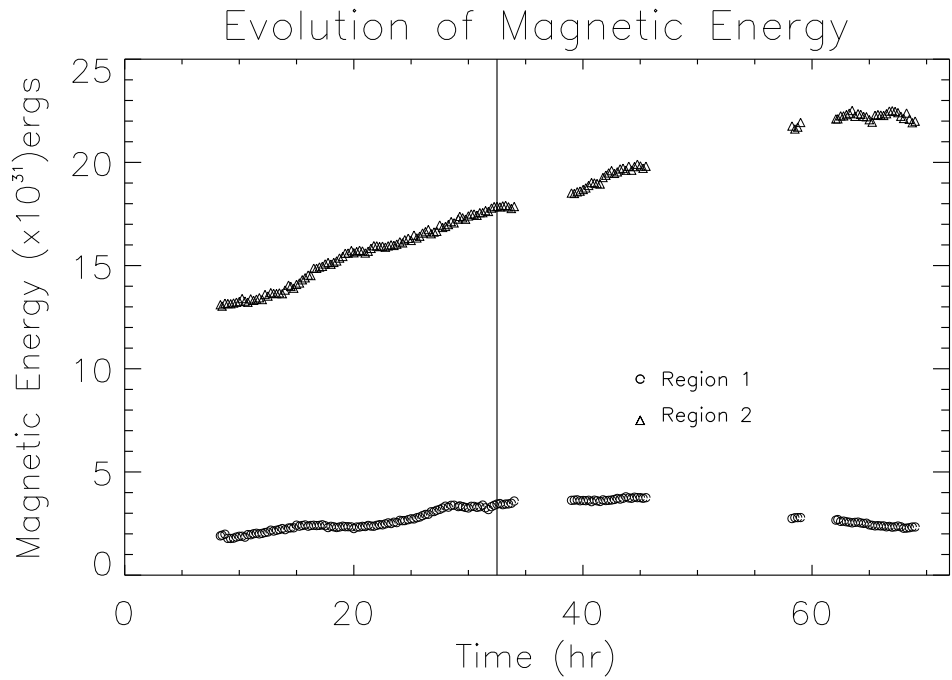}
\caption{This plot shows the evolution of the magnetic potential energy
(measured in ergs) with time, in the separate regions marked by 1,2.}
\label{fig6}
\end{figure}

\begin{figure}
\includegraphics[width=15cm]{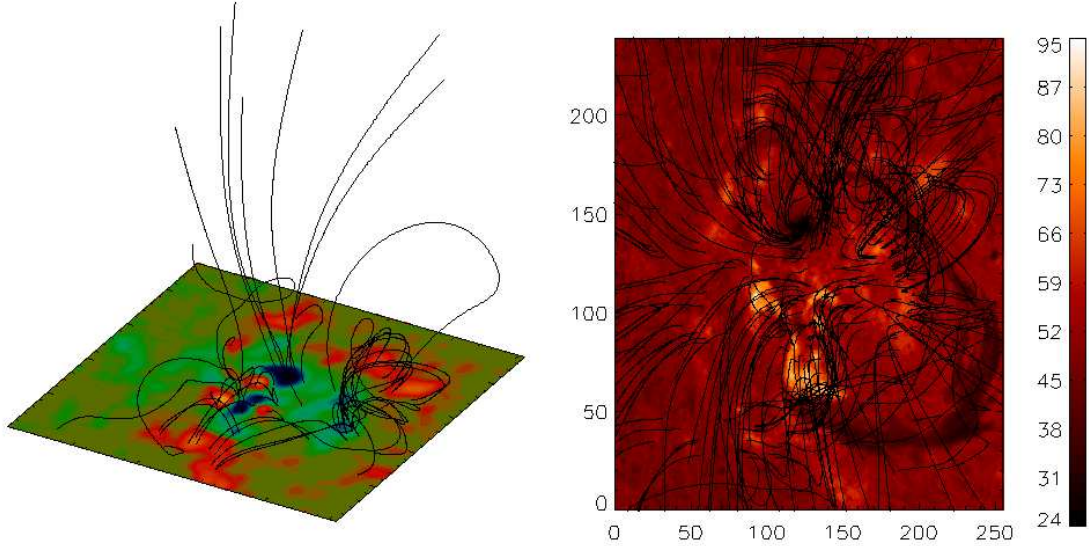}
\caption{The above figure shows the nonlinear extrapolated magnetic field lines in the NOAA AR10501
from the vector magnetogram data obtained  on 18 November at 00:20
UT from the Solar flare telescope at Mitaka.  For this purpose, we
considered the entire box of $256 \times 240\times 240$ pixels,
where 1 pixel =1.32 arc-sec and used the technique for field
extrapolation  developed by Wiegelmann (2004). The non-linear
force free extrapolation field lines (shown in black) have been
overplotted in this figure. The size of the magnetogram is $256
\times 240$ pixels in x and y directions respectively. The field
has been extrapolated to 200 pixels in the z-direction. On the right is shown the projection of both open and closed field lines onto the H$_{\alpha}$ image}.
\label{fig7}
\end{figure}

%
%
%
%
%
% ...
%

%
%
\end{article}

\end{document}